\documentclass[sigconf]{acmart}
\renewcommand\footnotetextcopyrightpermission[1]{}
\usepackage{booktabs}

\setcopyright{none}
\acmConference[ACM SIGDA Ph.D. Forum 2019]{}
\acmYear{2019}

\acmArticle{4}
\acmPrice{15.00}

\begin{document}
\title{Collaborative Heterogeneous Computing on MPSoCs}
\subtitle{Extended Abstract}
\subtitlenote{Accepted to ACM SIGDA Ph.D. Forum at Design Automation Conference (DAC) 2019. Siqi Wang is with the Department of Computer Science, School of Computing, National University of Singapore, SG. 
	E-mail:~(wangsq@comp.nus.edu.sg)}

\author{Siqi Wang}
\affiliation{%
	\city{} 
}

\maketitle

\section{Introduction}
With the emerging demand for computations on mobile devices, heterogeneous multi-processor system-on-chips~(MPSoCs) are envisioned to dominate the current and future mobile computing landscape. Heterogeneous MPSoCs usually comprise of various processing elements 
such as general-purpose cores (CPUs) with different performance-power characteristics and application-specific accelerators, examples of which are graphics processing units (GPUs), digital signal processors (DSPs), reconfigurable accelerators (FPGAs, etc.) and the recent neural acceleration engines (NPUs, etc.). Such heterogeneity presented on the SoC enables delicate matching of computational kernels to the processing elements that are best suited to perform the computation, which leads to substantial improvements in performance and energy-efficiency.

The heterogeneity can be broadly classified into performance and functional heterogeneity, while commercial SoCs are trending toward adopting both in the same chip. {Performance heterogeneity} consists of cores with the same functionality (instruction-set architecture, ISA) but with different power-performance characteristics, an example of which is the ARM big.LITTLE CPU architecture. The difference stems from distinct micro-architectural features such as in-order core versus out-of-order core. The complex cores provide better performance at the cost of higher power consumption while the simpler cores exhibit low-power behavior with lower performance. 
{Functional heterogeneity} features cores with very different functionality (different ISA) existing on the same die. The heterogeneity takes advantage of certain execution pattern for exceptional speed-up to meet the performance requirement under the stringent power budget. 
Under carefully managed exploitation of multiple forms of heterogeneity, heterogeneous MPSoCs present great potential to sustain the performance and power requirements for next generation mobile computing.

While architectural heterogeneity is promising, software development efforts are required to fully benefit from this architectural advancement~\cite{isvlsi}. This thesis (extended abstract) presents the software development efforts toward efficient exploitation of heterogeneity through intricate mapping of computational kernels, collaborative execution of multiple processing elements and application specific techniques. The goal is to embrace the heterogeneity to unleash the full potential of the heterogeneous MPSoCs towards high-performance energy-efficient mobile computing.

\section{Exploitation of Heterogeneity}

Functional heterogeneity presents application developers with a diverse choice of processing elements on the same chip. They now have the opportunity and the responsibility to take advantage of the unique characteristics of different processing elements to improve execution performance. However, the matching of computational kernels to processing elements is difficult as the performance is a complex interplay among the exposed parallelism, the compiler, and the processor architecture.
Furthermore, the application kernel needs to be implemented in different processor-specific languages to measure the performance of each processing element. If the performance of the applications on different processing elements are made available at an early stage, the developers will then be able to make an informed decision in selecting the most appropriate processing element and concentrate on further processor-specific languages and optimizations.
{\em CGPredict}~\cite{cgpredict}  is proposed to guide developers in the early design choice without tedious redevelopment efforts. It is an analytical framework that accurately estimates the performance of a computational kernel on an embedded GPU architecture from unoptimized, single-threaded C code.

CGPredict takes a computational kernel in the form of single-threaded C code and generates its execution trace through a \textit{Trace Extraction} phase. In order to emulate the behavior of GPU, a \textit{Warp Formation} phase is introduced to transform the single-threaded trace into its multi-threaded equivalent. CGPredict then extracts computation (compute instructions) and memory access information. The compute cycle count is obtained by mapping compute instructions to GPU instructions in the \textit{Computation Analysis} stage, while the memory cycle count is obtained through memory access information analysis with access patterns and cache behavior in the \textit{Memory Behavior Analysis} stage. The results from the two analysis stages complete the execution characteristics we need from the kernel for performance prediction. Lastly, together with the hardware architectural parameters obtained from micro- benchmarking, a comprehensive \textit{Analytical Prediction Model} is engaged to predict the final execution performance using the computation and memory execution characteristics.

CGPredict provides accurate GPU performance estimations from only C code with 9\% error.
It also provides insights regarding the characteristics of the kernel and the GPU that influence performance, such as coalescing of memory accesses and shared memory usage. These insights offer opportunities for the developers to understand the intrinsic strengths and weaknesses of the architecture in the context of a particular kernel that can facilitate further code optimizations. Furthermore, CGPredict in conjunction with an existing FPGA performance predictor from C code~\cite{fgpa} achieves our objective of making the perfect choice of processing elements (CPU, GPU or FPGA) given a kernel.

\section{Co-execution on Mobile Platform}
The ever-increasing processing requirements impose higher pressure on mobile devices with limited processing capability. Executing an application on a single processing element may not sustain the performance requirements, while other processing elements that can potentially be used are not actively contributing. The concurrent co-execution of a single computational kernel on multiple processing elements thus exhibits great potential in achieving additional performance. The design space of co-execution is huge with the exploitation of both performance and functional heterogeneity. In addition, the ability to vary clock frequencies enables the compromise between the achievable performance and power consumption which further extends the design space. We show through exhaustive design space search~\cite{iccd} that by executing a computational kernel simultaneously on all available processing elements (big.LITTLE CPU cores, GPUs) together with suitable voltage-frequency settings for all these cores, as high as 39\% energy savings and 19\% improvement in runtime are achieved compared to the stand-alone executions. The improvement in runtime allows developers to have more flexibility in tuning the various voltage-frequency settings to achieve higher performance with certain constraints.

On the other hand, the inherent characteristics of mobile systems demand stringent power and thermal requirements as compared to server system; this is especially so because of the lack of active cooling measures on mobile devices.
Commercial heterogeneous MPSoCs usually implement operating system level thermal management techniques such as processor frequency throttling to prevent failure of the chip at high temperatures. Engaging multiple processing elements concurrently may expedite the heating up of the system, necessitating frequency throttling and hence degradation of performance. Therefore, the benefit of co-execution can be compromised by the throttling of frequency due to thermal issues. We propose \textit{OPTiC}~\cite{optic} to anticipate such thermal impact on execution when engaging multiple processing elements for performance optimization. 

OPTiC presents a static partitioning strategy to split a computational kernel across CPU and GPU cores for concurrent execution, with the voltage-frequency settings of the cores carefully determined considering the thermal effects. OPTiC builds on extensive and comprehensive modeling of power and runtime, resource contention and thermal behavior. The power and runtime of the CPU and GPU cores at all frequencies are predicted through analytical modeling from one profile run at a sample frequency. The thermal behavior is captured through a \textit{thermal throttling model} that predicts the occurrence of OS frequency throttling and the resultant runtime under such thermal condition. From the individual performances, the allocation of the workload and the co-execution performance are predicted through a \textit{co-execution model} that considers the effect of thermal frequency throttling and resource contention. The framework then goes through all the possible frequency settings and predicts the performance to locate the optimal configuration and workload allocation. While the performance of an application is largely affected by thermal conditions, OPTiC is able to predict the configuration that presents on average 14\% runtime improvement over standalone execution. OPTiC further demonstrates great temperature control with real-life applications. With the configuration predicted by OPTiC, the chip exhibits a much cooler temperature as compared to the Linux frequency governors.

\section{Toward machine learning}
Lastly, the rise of machine learning applications poses great challenges to mobile platforms.
Deploying neural network inferencing on mobile platforms require the exploitation of heterogeneity to sustain the performance requirements given limited resources and stringent power budgets.
Although dedicated neural accelerators (NPUs, etc.) show exceptional speed-ups for applications like convolutional neural network (CNN), the technique is highly platform dependent and not applicable to general architectures without the accelerator. Furthermore, CNN are more commonly used as building blocks to construct more complex systems. We envision in the near future that multiple independent inference sub-tasks are expected to be performed concurrently. This requires all the available processing elements to run the inference engines in parallel. Therefore, it is important to develop general techniques that are applicable to existing heterogeneous MPSoCs on mobile platforms. 

Commercial CNN libraries usually only engage one of the processing elements and are often ignorant to the co-execution of multiple processing elements. ARM Compute Library (ARM-CL) provides out-of-the-box support for parallel execution through multi-threading for the CPU clusters. But the concurrent co-execution of the big and LITTLE cluster with multi-threading is harmful for performance due to cache coherence overheads. Thus, the kernel-level splitting among processing elements fails to either reduce the end-to-end latency or the throughput. We present an alternative framework \textit{pipe-it}~\cite{pipeit} that employs a pipelined design to split the convolutional layers across processing elements (different CPU clusters) to improve throughput for streaming inferencing. Here, the two CPU core clusters are divided into multiple sub-core-clusters as processing elements to construct the pipeline stages to better match the resources and workload. Pipe-it includes an analytical performance model that predicts the performance of a convolutional layer on different configurations (core type, count) from its network structure descriptions. The predicted performance is then used as input into a design space exploration algorithm that navigates the design space and locates the best fitting pipeline configuration and respective layer allocation. Pipe-it with the predicted multi-stage pipeline achieves on average 39\% throughput gain compared with the execution on a single processing element.

\bibliographystyle{ACM-Reference-Format}
\bibliography{ref} 

\end{document}